\documentclass{article}
\usepackage{latexsym, a4wide}
\usepackage{amsmath, rotating, color}
\usepackage{amsfonts,amssymb, amsthm, mathrsfs}
\usepackage{bbm}
\usepackage{wrapfig}
\usepackage{amsmath}
\usepackage{natbib}
\usepackage[intlimits]{empheq}
\usepackage{multicol} 
\usepackage{amsthm}
\newcommand\unnumberedfootnote[1]{ %
        \let\temp=\thefootnote %
        \renewcommand{\thefootnote}{}%
        \footnote{#1}%
        \let\thefootnote=\temp%
        \addtocounter{footnote}{-1}}

\newtheorem{theorem}{Theorem}
\newtheorem{proposition}{Proposition}[section]

\theoremstyle{definition}
\newtheorem{remark}[proposition]{Remark}

\numberwithin{equation}{section}

\DeclareMathAlphabet{\mathpzc}{OT1}{pzc}{m}{it}

\newcommand{\eins}{\setlength{\unitlength}{0,2cm}
  \begin{picture}(2,3)(0,1) \put (1,0){\line(0,1){1}} \put
    (1,2){\line(0,1){1}} \put (1,1.2){\line(0,1){.1}} \put
    (1,1.4){\line(0,1){.1}} \put (1,1.6){\line(0,1){.1}} \put
    (1,1.8){\line(0,1){.1}}
  \end{picture}
}

\newcommand{\zwei}{
  \setlength{\unitlength}{0,2cm}
  \begin{picture}(2,3)(-.5,1) \put (0,3){\line(0,-1){3}} \put
    (0,2){\line(1,0){1}} \put (1,2){\line(0,-1){2}}
  \end{picture}
}

\newcommand{\dreinocoal}{    
  \setlength{\unitlength}{0,2cm} 
  \begin{picture}(4,4)(-1,1) 
    \put (1,3){\line(0,-1){1.5}} 
    \put (0,2){\line(1,0){2}} 
    \put (0,2){\line(0,-1){2}} 
    \put (1,2){\line(0,-1){2}} 
    \put (2,2){\line(0,-1){2}}
    \put (0,2.2){\makebox(0,0){\tiny c}}
    \put (2,2.2){\makebox(0,0){\tiny i}}
  \end{picture}
}

\newcommand{\dreionecoala}{
\setlength{\unitlength}{0,2cm}
  \begin{picture}(4,4)(-1,1) \put (1,3){\line(0,-1){1}} \put
    (0,2){\line(1,0){2}} \put (0,2){\line(0,-1){1}} \put
    (1,2){\line(0,-1){1}} \put (2,2){\line(0,-1){2}} \put
    (0,1){\line(1,0){1}} \put (0.5,1){\line(0,-1){1}} \put
    (0,2.2){\makebox(0,0){\tiny c}} \put (2,2.2){\makebox(0,0){\tiny
        i}}
  \end{picture}
}

\newcommand{\dreionecoalb}{    
  \setlength{\unitlength}{0,2cm}
  \begin{picture}(4,4)(-1,1) \put (1,3){\line(0,-1){1}} \put
    (0,2){\line(1,0){2}} \put (0,2){\line(0,-1){2}} \put
    (1,2){\line(0,-1){1}} \put (2,2){\line(0,-1){1}} \put
    (1,1){\line(1,0){1}} \put (1.5,1){\line(0,-1){1}}
    \put (0,2.2){\makebox(0,0){\tiny c}}
    \put (2,2.2){\makebox(0,0){\tiny i}}
  \end{picture}
}

\newcommand{\dreionecoalc}{
  \setlength{\unitlength}{0,2cm}
  \begin{picture}(4,4)(-1,1) \put (0,3){\line(0,-1){3}} \put
    (0,2){\line(1,0){2}} \put (1,2){\line(0,-1){2}} \put
    (2,2){\line(0,-1){1}} \put (2,1){\line(-1,0){1}}
    \put (1,2.2){\makebox(0,0){\tiny c}}
    \put (2,2.2){\makebox(0,0){\tiny i}}
  \end{picture}
}

\begin{document}
\title{\LARGE Populations under frequency-dependent selection and
  genetic drift: a genealogical approach}

\thispagestyle{empty}

\author{{\sc by P. Pfaffelhuber and B. Vogt} \\[0ex]
  \emph{Albert-Ludwigs-Universit\"at Freiburg}}
\date{\today}


\maketitle
\unnumberedfootnote{\emph{AMS 2000 subject classification.} {\tt
    91A22} (Primary) {\tt 60K35, 92D15, 91A15} (Secondary).}

\unnumberedfootnote{\emph{Keywords and phrases.} Evolutionary game
  theory, weak selection, ancestral selection graph}

\begin{abstract}
\noindent
Evolutionary models for populations of constant size are frequently
studied using the Moran model, the Wright--Fisher model, or their
diffusion limits. When evolution is neutral, a random genealogy given
through Kingman's coalescent is used in order to understand basic
properties of such models. Here, we address the use of a genealogical
perspective for models with weak frequency-dependent selection, i.e.\
$Ns =: \alpha$ is small, where $s$ is the fitness advantage of a fit
individual and $N$ is the population size. When computing fixation
probabilities, this leads either to the approach proposed by
\cite{Rousset2003}, who argues how to use Kingman's coalescent in
order to study weak selection, or to extensions of the ancestral
selection graph of \cite{NeuhauserKrone1997} and
\cite{Neuhauser1999}. As an application of this genealogical approach,
we re-derive the one-third rule of evolutionary game theory
\citep{Nowak2004}. In addition, we provide the approximate
distribution of the genealogical distance of two randomly sampled
individuals under linear frequency-dependence.
\end{abstract}

\section{Introduction}
Evolutionary game theory has long been studied using replicator
equations and deterministic systems by using an infinite population
limit \citep{TaylorJonker1978, Maynard-Smith1982,
  Nowak2006}. Recently, the amount of research dealing with repeated
games in finite populations increases. In the simplest case, a
population of~$N$ players (each carrying one out of a possible set of
strategies) repeatedly chooses an opponent at random and plays a game
with payoff matrix $M$. In addition, players can produce offspring and
the fitness of an individual with strategy~$A$ is determined by the
average payoff the player obtains in the evolutionary game. Since the
average payoff depends on the frequencies of the strategies in the
population, the fitness is frequency-dependent, and the frequency path
is e.g.\ given by a Moran model or a Wright--Fisher model with
frequency-dependent selection.

When studying models for populations of individuals carrying different
types (or adopting different strategies), several questions are most
important (see e.g.\ \citealp{Ewens2004}). What is the probability of
fixation of a type (in absence of mutation), or what is the random
distribution of types in mutation-selection balance? If fixation
occurs, what is the distribution of the fixation time? When taking a
sample from the population, what is its genealogical structure? In
addition, extensions to structured populations or to the multi-type
case are considered. For the questions on fixation probabilities and
times, the theory of birth-death processes and (one-dimensional)
diffusions is frequently used \citep{KarlinTaylor1981}. Here, explicit
formulas are available for such quantities, but these can hardly be
extended to multi-dimensional cases. Another approach is the use of
the genealogical structure by coalescent processes
\citep{Kingman1982a, Berestycki2009}. The advantage is that it is
extensible to populations in structured or multi-type populations. In
addition, questions on the genealogical structure can be asked. As an
example from evolutionary game theory, \cite{GokhaleTraulsen2011}
study the frequency path of many-player many-strategy games using such
a genealogical approach.

The current paper was motivated by the \emph{one-third rule} of
evolutionary game theory \citep{Nowak2004}. It states that in finite
populations of players with two strategies $A$ and $B$, strategy $A$
is favored in the sense that the fixation probability of type $A$ in a
$B$-population is higher than under neutrality if and only if type~$A$
is favored in the infinite population limit at frequency 1/3. Since
the one-third rule was found in a Moran model, it has as well been
proven to hold in a Wright--Fisher model \citep{Traulsen2006,
  Imhof2006} and in general exchangeable models
\citep{Lessard2007}. An intuitive explanation was given in
\cite{Ohtsuki2007}, who argue that an individual interacts on average
with $B$-players twice as often as with $A$-players, which implies the
one-third rule. \cite{WuTraulsen2010} discuss violations of the
one-third rule for non-linear frequency dependence.

~

The goal of the present paper is to use a genealogical approach in the
study of fixation probabilities. Only \cite{Ladret2007}, who use a
method developed by \cite{Rousset2003}, already take this route; see
also~\cite{Rousset2004}, p.\ 92f. We complement their approach by
using the Ancestral Selection Graph (ASG), first introduced in the
frequency-independent case by~\cite{NeuhauserKrone1997} and
\cite{KroneNeuhauser1997}. This graph extends the coalescent to the
selective case and has been further extended to a simple
frequency-dependent case in \cite{Neuhauser1999}. In addition, we
study genealogical distances in models from evolutionary game theory,
i.e.\ frequency-dependent selection.

~

The paper is organized as follows: After stating the models (and their
diffusion limits), we re-formulate the one-third rule and our results
on genealogical distances. Then, we provide two proofs of the
one-third rule. Finally our genealogical result is proved.

\section{Models}
We are studying the solution $\mathcal X = (X_t)_{t\geq 0}$ of the
stochastic differential equation
\begin{align}\label{eq:SDE}
  d X & = ((1-X)\theta_A - X\theta_B)dt + \alpha X(1-X)(\beta - \gamma
  X)dt + \sqrt{X(1-X)}dW
\end{align}
for $\alpha, \theta_A, \theta_B\geq 0, \beta, \gamma\in\mathbb R$.
Here, $\alpha$ is called the \emph{scaled selection coefficient}, and
$\theta_A$ and $\theta_B$ the \emph{scaled mutation rates}, and
$\mathcal X$ gives the frequency path of type (or strategy) $A$ in a
population of constant size. For this diffusion, we define in the case
$\theta_A = \theta_B=0$
$$ T_1 := \inf\{t\geq 0: X_t=1\}$$
and note that $T_1<\infty$ if and only if type-$A$ eventually fixes in
the population. In our analysis, the version of~\eqref{eq:SDE} without
mutation and without fluctuations, 
\begin{align}
  \label{eq:ODE}
  dx = \alpha x(1-x)(\beta - \gamma x)dt
\end{align}
will also play a role.

\subsection{A finite Moran model}
\label{ss:moran}
We are turning to the frequency path of strategy (or type) $A$ in a
time-continuous Moran model of size $N$, which converges to $\mathcal
X$ as $N\to\infty$. We distinguish four cases, corresponding to the
signs of $\beta$ and $\gamma$.  Setting $x_\ast = \beta/\gamma$ (which
is an equilibrium point in~\eqref{eq:ODE}), the four cases are:
\begin{enumerate}
\item[(i)] $\beta>0$ and $\gamma>0$, which leads to a stable fixed
  point in $x_\ast$ for~\eqref{eq:ODE} (if $x_\ast\in(0,1)$),
\item[(ii)] $\beta<0$ and $\gamma<0$, which leads to an unstable fixed
  point in $x_\ast$  for~\eqref{eq:ODE} (if $x_\ast\in(0,1)$),
\item[(iii)] $\beta>0$ and $\gamma<0$, which leads $x=1$ as the only
  stable fixed point in~\eqref{eq:ODE},
\item[(iv)] $\beta<0$ and $\gamma>0$, which leads $x=0$ as the only
  stable fixed point in~\eqref{eq:ODE}.
\end{enumerate}
In case (ii) and (iv), strategy $A$ cannot invade a population
consisting only of strategy $B$ individuals in \eqref{eq:ODE} and vice
versa.

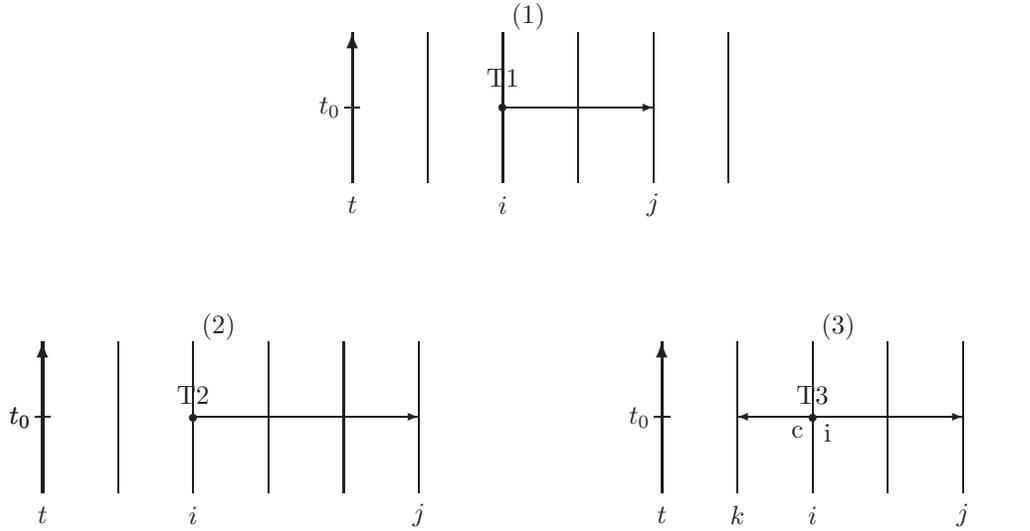
\begin{figure}
  \begin{center}
    \parbox{6cm}{
      \hspace{2cm} (1)\\
      \setlength{\unitlength}{1cm}
      \begin{picture}(5,2)(0,0)
        \put (0,0){\thicklines \vector(0,1){2}}
        \put (2,1){\vector(1,0){2}}
        \put (1,0){\line(0,1){2}}
        \put (2,0){\line(0,1){2}}
        \put (3,0){\line(0,1){2}}
        \put (4,0){\line(0,1){2}}
        \put (5,0){\line(0,1){2}}
        \put (-0.1,1){\line(1,0){0.2}}
        \put (2,1){\circle*{0,1}}
        \put (0, -0.3){\makebox(0,0){$t$}}
        \put (2, -0.3){\makebox(0,0){$i$}}
        \put (4, -0.3){\makebox(0,0){$j$}}
        \put (2, 1.4){\makebox(0,0){T1}}
        \put (-0.3, 1){\makebox(0,0){$t_0$}}
      \end{picture}
    }
  \end{center}

  \vspace{1cm}
  
  \begin{center}
    \parbox{6cm}{
      \hspace{2cm} (2)\\
      \setlength{\unitlength}{1cm}
      \begin{picture}(5,2)(0,0)
        \put (0,0){\thicklines \vector(0,1){2}}
        \put (-0.3, 1){\makebox(0,0){$t_0$}}
        \put (2,1){\vector(1,0){3}}
        \put (1,0){\line(0,1){2}}
        \put (2,0){\line(0,1){2}}
        \put (3,0){\line(0,1){2}}
        \put (4,0){\line(0,1){2}}
        \put (5,0){\line(0,1){2}}
        \put (-0.1,1){\line(1,0){0.2}}
        \put (2,1){\circle*{0,1}}
        \put (0, -0.3){\makebox(0,0){$t$}}
        \put (2, -0.3){\makebox(0,0){$i$}}
        \put (5, -0.3){\makebox(0,0){$j$}}
        \put (2, 1.3){\makebox(0,0){T2}}
        \put (-0.3, 1){\makebox(0,0){$t_0$}}
      \end{picture}
    } \hspace{2cm}
    \parbox{6cm}{
      \hspace{2cm} (3)\\
      \setlength{\unitlength}{1cm}
      \begin{picture}(5,2)(0,0)
        \put (0,0){\thicklines \vector(0,1){2}}
        \put (2,1){\vector(1,0){2}}
        \put (2,1){\vector(-1,0){1}}
        \put (1,0){\line(0,1){2}}
        \put (2,0){\line(0,1){2}}
        \put (3,0){\line(0,1){2}}
        \put (4,0){\line(0,1){2}}
        \put (5,0){\line(0,1){2}}
        \put (-0.1,1){\line(1,0){0.2}}
        \put (2,1){\circle*{0,1}}
        \put (0, -0.3){\makebox(0,0){$t$}}
        \put (1.8,0.8){\makebox(0,0){c}}
        \put (2.2,0.8){\makebox(0,0){i}}
        \put (2, -0.3){\makebox(0,0){$i$}}
        \put (4, -0.3){\makebox(0,0){$j$}}
        \put (1, -0.3){\makebox(0,0){$k$}}
        \put (2, 1.3){\makebox(0,0){T3}}
        \put (-0.3, 1){\makebox(0,0){$t_0$}}
      \end{picture}
    }
  \end{center}
  \caption{\label{fig1}Three different kinds of events occur in the
    Moran model which has limit~\eqref{eq:SDE}. (1) For every
    (ordered) pair of individuals $i$ and $j$, a T1-arrow occurs at
    rate $\tfrac 12$. Here, the offspring of $i$ replaces individual
    $j$ (independent of the types of $i$ and $j$).  (2) T2-arrows are
    the first of two selective arrows. In cases~(i) and (iii)
    (cases~(ii) and~(iv)), they occur at rate $\alpha\beta$ (at rate
    $-\alpha\beta$). Upon such an event, the offspring of
    individual~$i$ replaces $j$, given that $i$ has type $A$ (type
    $B$).  (3)~T3-(double-)arrows, occur in case~(i) and (iv)
    (cases~(ii) and~(iii)) at rate $\alpha\gamma$ (at rate~$-\alpha
    \gamma$). Here, the imitating arrow is created from~$i$ to $j$, as
    well as the checking arrow from~$i$ to~$k$. Given $i$ has type~$B$
    (type~$A$) the checked individual~$k$ has type~$A$ (in both
    cases), individual~$j$ is replaced by an offspring of
    individual~$i$.  }
\end{figure}

\begin{figure}
  \centering
  \setlength{\unitlength}{0.7cm}
  \begin{picture}(6,10.5)(0,-0.5)
    \put (0,.5){\thicklines \vector(0,1){10}}
    \put (1,10.5){\line(0,-1){10}}
    \put (2,10.5){\line(0,-1){10}}
    \put (3,10.5){\line(0,-1){10}}
    \put (4,10.5){\line(0,-1){10}}
    \put (5,10.5){\line(0,-1){10}}
    \put (6,10.5){\line(0,-1){10}}
    \put (4,9.5){\vector(-1,0){1}}
    \put (4,9.5){\circle*{0,1}}
    \put (4,9.7){\makebox(0,0){\begin{tiny} T1 \end{tiny}}}
    
    \put (3,7.7){\vector(1,0){1} }
    \put (3,7.7){\vector(-1,0){2}}
    \put (3,7.7){\circle*{0,1}}
    \put (3,7.9){\makebox(0,0){\begin{tiny} T3 \end{tiny}}}
    
    \put (4,4.5){\vector(-1,0){2}}
    \put (4,4.5){\circle*{0,1}}
    \put (4,4.7){\makebox(0,0){\begin{tiny} T1 \end{tiny}}}
  
    \put (2,3.5){\vector(1,0){1}}
    \put (2,3.5){\circle*{0,1}}
    \put (2,3.7){\makebox(0,0){\begin{tiny} T2 \end{tiny}}}
    
    \put (1,2.3){\vector(1,0){1}}
    \put (1,2.3){\circle*{0,1}}
    \put (1,2.5){\makebox(0,0){\begin{tiny} T1 \end{tiny}}}
    
    \put (3,2.1){\vector(1,0){2}}
    \put (3,2.1){\circle*{0,1}}
    \put (3,2.3){\makebox(0,0){\begin{tiny} T1 \end{tiny}}}
    
    \put (5,1){\vector(-1,0){3}}
    \put (5,1){\circle*{0,1}}
    \put (5,1.2){\makebox(0,0){\begin{tiny} T1 \end{tiny}}}
    
    \put (6,8.5){\vector(-1,0){2}}
    \put (6,8.5){\circle*{0,1}}
    \put (6,8.7){\makebox(0,0){\begin{tiny} T2 \end{tiny}}}
    
    \put (5,6.5){\vector(-1,0){2}}
    \put (5,6.5){\circle*{0,1}}
    \put (5,6.7){\makebox(0,0){\begin{tiny} T1 \end{tiny}}}
    
    \put (5,4.1){\vector(-1,0){1}}
    \put (5,4.1){\vector(1,0){1}}
    \put (5,4.1){\circle*{0,1}}
    \put (5,4.3){\makebox(0,0){\begin{tiny} T3 \end{tiny}}}
    
    \put (0, 10.8){\makebox(0,0){$t$}}
    \put (4.5, 4.2){\makebox(0,0){\begin{tiny}i\end{tiny}}}
    \put (5.5, 4.2){\makebox(0,0){\begin{tiny}c\end{tiny}}}
    \put (3.5, 7.8){\makebox(0,0){\begin{tiny}i\end{tiny}}}
    \put (2.5, 7.8){\makebox(0,0){\begin{tiny}c\end{tiny}}}
  \end{picture}
  \caption{\label{fig2} The graphical representation of the Moran
    model (without mutation). Here, every pair of individuals is
    connected by a T1-arrow at rate~1, every individual is the origin
    of a T2-arrow at rate $\alpha|\beta|$, and every individual is
    origin of a T3-double-arrow to one checking and one imitating line
    at rate $\alpha|\gamma|$. Types/strategies are inherited along the
    arrows according to the rules 1.--3.}
\end{figure}
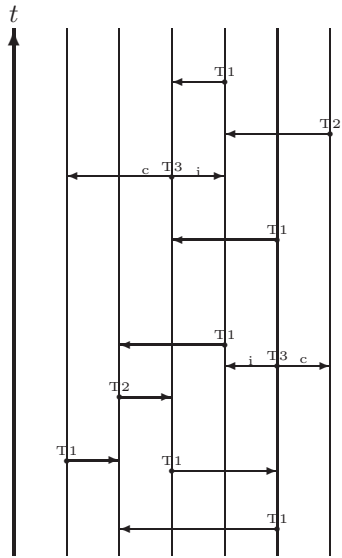

The time-continuous Moran model is a Markov process with state space
$\{A,B\}^N$ (indicating the types of $N$ individuals) and the
following transition rules:
\begin{enumerate}
\item Every pair of individuals $i$ and $j$ resamples at rate~1. Upon
  a resampling event, the offspring of $i$ (or $j$) replaces $j$ (or
  $i$) with probability $\tfrac 12$.
\item[2A.] In cases (i) and (iii), i.e.\ when $\beta>0$, the offspring of
  any $A$-individual replaces a randomly chosen individual at rate
  $\alpha\beta$.
\item[2B.] In cases (ii) and (iv), i.e.\ when $\beta<0$, the offspring
  of any $B$-individual replaces a randomly chosen individual at rate
  $-\alpha\beta = \alpha|\beta|$.
\item[3A.] In cases (i) and (iv), i.e.\ when $\gamma>0$, every
  $B$-individual picks a randomly chosen individual at rate
  $\alpha\gamma$. If it is an $A$-individual, an offspring of the
  $B$-individual replaces another, randomly chosen individual.
\item[3B.] In cases (ii) and (iii), i.e.\ when $\gamma<0$, every
  $A$-individual picks a randomly chosen individual at rate
  $-\alpha\gamma = \alpha|\gamma|$. If it is another $A$-individual,
  an offspring of the first $A$-individual replaces another,
  randomly chosen individual.
\item[4.] Every individual is hit by a mutation event at rate
  $\bar\theta := \theta_A + \theta_B$. With probability
  $\theta_A/\bar\theta$, the individual turns into type $A$, and with
  probability $\theta_B/\bar\theta$, it turns to $B$, independent of
  the type of the parent.
\end{enumerate}
In the graphical construction of the Moran model, the transitions
1.--3.\ are given by arrows denoted T1-T3, respectively.  The
T3-arrows are in fact double-arrows since the type of one individual
has to be checked (along the line which we will call the checking
line; see also~\cite{Neuhauser1999}) while the second arrow leads to
reproduction (which will be called the imitating line). See
Figures~\ref{fig1} and~\ref{fig2} for illustrations of the transitions
1.,\ 2.\ and 3. By now, it is a classical result that the frequency
path of the Moran model converges weakly (with respect to the topology
of uniform convergence on compact sets) to solutions
of~\eqref{eq:SDE}; see e.g.\ \cite{EthierKurtz1993}.

\subsection{Connection with evolutionary games}
Consider a game with payoff matrix 
\begin{center}
  \begin{tabular}{c|cc}
    & $A$ & $B$ \\\hline
    $A$ & $a$ & $b$ \\
    $B$ & $c$ & $d$
  \end{tabular}
\end{center}
There are $N$ players, each one adopting either strategy (or type) $A$
or $B$. If the frequencies of type $A$ and type $B$ are $X_A$ and
$X_B$, respectively, the fitnesses are dependent on the payoff matrix
and are given by
\begin{equation}
  \label{eq:fAB}
  \begin{aligned}
    f_A &:= \tfrac 12 + s(aX_A + bX_B) = \tfrac 12 + s(b + (a-b)X_A),\\
    f_B &:= \tfrac 12 + s(cX_A + dX_B) = \tfrac 12 + s(d + (c-d)X_A).
  \end{aligned}
\end{equation}
Note that $aX_A+bX_B$ is the average payoff of a type-$A$ individual
when playing against a random player from the population. 

Here, the offspring of every individual of type $A$ replaces a
randomly chosen individual at rate $f_A$. Accordingly, a type-$B$
offspring replaces a randomly chosen individual at rate $f_B$ (given
that $s$ is small enough such that the rates are non-negative). The
frequency path of type-$A$, follows in the infinite-population limit,
$N\to\infty$,
\begin{align*}
  dx & = sx(1-x)(b+(a-b)x - d - (c-d)x)dt = sx(1-x)(\beta - \gamma
  x)dt
\end{align*}
with
\begin{align}
  \beta := b-d, \qquad \gamma := b-d+c-a.
\end{align}
Moreover, when $N\to\infty$, $s\to 0$ such that $Ns\to \alpha$, and
time is rescaled by a factor of $N$, the frequency path of type-$A$
follows the SDE as given in~\eqref{eq:SDE}. Since the Moran model from
Section~\ref{ss:moran} has the same diffusion limit, we use its
graphical representation in order to obtain insights into the
evolutionary game just described.

~

\section{Results}
Starting by re-formulating the \emph{one-third rule} in
Section~\ref{S:onethird}, we give our result on the distribution of
genealogical distances in the Moran model from Section~\ref{ss:moran}
in Section~\ref{ss:dist}.

\subsection{The one-third rule}
\label{S:onethird}
We are now ready to formulate the one-third rule. For
$\theta_A=\theta_B=0$, we define
$$ p_{\text{fix}}(\varepsilon,\alpha) :=  \mathbf P(T_1 < \infty| X_0=\varepsilon),$$
which is the fixation probability of type $A$ if it starts in
frequency $\varepsilon$ and selection intensity is $\alpha$. Note that
$p_{\text{fix}}(\varepsilon,0) = \varepsilon$, i.e.\ the chance that
strategy $A$ fixes equals its starting frequency, if evolution is
neutral.

\begin{theorem}[One-third rule, \cite{Nowak2004}]
  \label{T:onethird}
  Let $\mathcal X = (X_t)_{t\geq 0}$ be the solution of~\eqref{eq:SDE}
  with $\theta_A = \theta_B = 0$. Then,
  \begin{align*}
    \lim_{\alpha\to 0} \frac 1
    {\alpha\varepsilon}\big(p_{\text{fix}}(\varepsilon,\alpha) -
    \varepsilon\big) \xrightarrow{\varepsilon\to 0} \beta - \gamma
    \frac 13 .
  \end{align*}
  In particular, $p_{\text{fix}}(\varepsilon,\alpha) >
  p_{\text{fix}}(\varepsilon,0)$ for $\alpha, \varepsilon$ small
  enough, if and only if
  $$ \beta - \gamma \frac 13 > 0,$$
  i.e.\ the fitness of type $A$ is above average at frequency $x=\frac
  13$.
\end{theorem}

\noindent
In Section~\ref{S:twop}, we will provide two different proofs of this
result. The first proof goes back to~\cite{Ladret2007}, the second one
uses the ancestral selection graph which we introduce in
Section~\ref{ss:ASG}.

\subsection{Genealogical distances}
\label{ss:dist}
In order to precisely formulate our results on genealogical distances,
we are considering the Moran model from Section~\ref{ss:moran}
including mutation, i.e.\ $\theta_A, \theta_B>0$. Assume that the
Moran model (which evolves according to the rules 1.--4.) has run for
an infinite amount of time. Then, every pair of individuals has a most
recent common ancestor which can be read off from the graphical
representation; see Figure~\ref{fig2}.

Here, we denote by $\mathbf P_{N,\alpha}$ the distribution of the
Moran model with population size $N$ and selection coefficient
$\alpha$ in equilibrium, and by $\mathbf E_{N,\alpha}$ the
corresponding expectation. Note that the Moran model is finite and
hence, the equilibrium exists and is unique. Moreover, we denote by
$R_{12}$ the genealogical distance of two randomly sampled individuals
from the equilibrium population. In the case $\alpha=0$, recall that
the distance of two randomly sampled individuals is exponentially
distributed, and hence, $\mathbf E_{N,\alpha=0}[e^{-\lambda R_{12}}]
\xrightarrow{N\to\infty} 1/(1+\lambda)$. The following result
generalizes this fact to small $\alpha$, i.e.\ weak selection.

\begin{theorem}[Genealogical distances under weak frequency-dependent
  selection]\mbox{}\\
  \label{T:dist}
  Let $\mathbf E_{N,\alpha}$ and $R_{12}$ be as above. Then, with
  $\bar\theta = \theta_A + \theta_B$,
  \begin{align*}
    \lim_{N\to\infty} \mathbf E_{N,\alpha} [e^{-\lambda R_{12}}] =
    \frac{1}{1+\lambda} - \alpha \gamma \frac{\theta_A\theta_B}{\bar
      \theta} \frac{(2+\bar\theta+\lambda)\lambda}{(6+\bar\theta +
      \lambda)(3+\bar\theta+\lambda)(1+\bar\theta +
      \lambda)(1+\bar\theta)(1+\lambda)^2} + \mathcal O(\alpha^2).
  \end{align*}
  In particular,
  \begin{align*}
    \lim_{N\to\infty} \mathbf E_{N,\alpha}[R_{12}] & = 1 + \alpha
    \gamma\frac{\theta_A\theta_B}{\bar\theta}
    \frac{2+\bar\theta}{(6+\bar\theta)(3+\bar\theta)(1+\bar\theta )^2}
    + \mathcal O(\alpha^2)
  \end{align*}
\end{theorem}

\begin{remark}
  \begin{enumerate}
  \item Note that -- up to first order in $\alpha$ -- the change in
    the genealogical distance only depends on $\gamma$ but not on
    $\beta$. This is not surprising since it has been shown that in
    the case of frequency-independent selection, $\beta=1, \gamma=0$,
    the change in the genealogical distance is of order $\alpha^2$;
    see Theorem 4.26 in \cite{KroneNeuhauser1997}. However, this leads
    to an interesting effect: assume that $\beta>\gamma>0$. Then, type
    $A$ is selectively advantageous at any time during the evolution
    of the population. So, one might guess that type-$A$ individuals
    have a higher chance to get offspring and genealogical distances
    are shorter than under neutrality. However, as the result shows,
    up to first order in $\alpha$, genealogical distances are larger
    than under neutrality. The reason is that for $\gamma>0$,
    selection is strongest by frequent interactions between~$A$
    and~$B$ individuals, and such interactions require a high
    heterozygosity, which in turn requires larger genealogical distance.
  \item In our proof, we implicitly take advantage of the recently
    developed theory of tree-valued stochastic processes from
    \cite{GrevenPfaffelhuberWinter2012} and \cite{DGP2012}. The idea
    is to describe the evolution of the genealogical distance of two
    randomly sampled points. If no events hit the two lines, the
    distance grows at constant speed. If a T1-arrow falls in between
    the two sampled individuals, their distance is reset to~0. Finally,
    with high probability, the two sampled lines are hit by T2- or
    T3-arrows only if the arrows originate from one of the $N-2$ other
    individuals. Once the evolution of the distance of two randomly
    sampled lines is given, we only have to find a(n approximate)
    fixed point in order to show Theorem~\ref{T:dist}. The details are
    given in Section~\ref{S:proof2}.
  \end{enumerate}
\end{remark}

\section{The ASG for linear frequency-dependent selection}
\label{ss:ASG}
The ancestral selection graph (ASG) was introduced by
\cite{NeuhauserKrone1997} and \cite{KroneNeuhauser1997} in order to
study genealogies in the case of frequency-independent selection.
Later, it was extended in \cite{Neuhauser1999} to a model of
minority-advantage, a special form of frequency-dependence in an
infinite alleles setting.

Here, we introduce the ASG in order to give a proof of
Theorem~\ref{T:onethird}. For its construction, consider again
Figure~\ref{fig2}. If a sample of individuals is drawn at the top of
the figure, one can read off all events which finally determine their
types. Because there are three different kinds of arrows, the history
of the sample as well comes with three different events. First,
T1-arrows between lines in the sample lead to coalescence events,
because common ancestors are found along such events. Second,
T2-arrows mostly origin from lines outside the sample. Since they as
well are determinants of the types in the sample, such events lead to
branching events into a continuing and an incoming branch. Third,
T3-arrows lead to splits of a line into the continuing, the checking
and the imitating line.

This informal description turns into the following stochastic process:
Starting with $n$ lines, the following transitions occur:
\begin{enumerate}
\item Every pair of lines coalesces at rate 1.
\item Every single line splits in two (called the continuing and
  incoming line) at rate $\alpha|\beta|$.
\item Every single line splits in three (called the continuing,
  checking and imitating line) at rate $\alpha|\gamma|$.
\end{enumerate}
Consider again the graphical representation of the Moran model from
Figure~\ref{fig2}. Here, starting with $n$ lines at the top of the
figure, it might be that the last event which hits one of the $n$
lines is a T2- or T3-arrow, which originates from one of the $n$
lines. However, since $N$ is assumed to be large, and the $N\to\infty$
limit gives~\eqref{eq:SDE}, this case can be ignored in the limit.

\begin{figure}
  \centering
  \setlength{\unitlength}{0.7cm}
 \hspace{1cm}  (A) \hspace{1.5cm} (B) \hspace{1.5cm} (C) \hspace{1.5cm} 
 (D) \hspace{1.5cm} (E) \hspace{1.5cm} (F)\\[2ex]
  \begin{picture}(0.5,3)(0,0)
    \put (1,0){\thicklines \vector(0,1){3}}
    \put (1, -0.3){\makebox(0,0){past}}
    \put (1, 3.3){\makebox(0,0){present}}
  \end{picture}
  \hspace{.5cm}
  \setlength{\unitlength}{0.7cm}
  \begin{picture}(2,3)(0,0)
    \put (1,0){\line(0,1){1}}
    \put (1,2){\line(0,1){1}}
    \put (1,1.2){\line(0,1){.1}}
    \put (1,1.4){\line(0,1){.1}}
    \put (1,1.6){\line(0,1){.1}}
    \put (1,1.8){\line(0,1){.1}}
  \end{picture}
  \hspace{.5cm}
  \setlength{\unitlength}{0.7cm}
  \begin{picture}(2,3)(0,0)
    \put (0,3){\line(0,-1){3}}
    \put (0,2){\line(1,0){1}}
    \put (1,2){\line(0,-1){2}}
  \end{picture}
  \hspace{.5cm}
  \setlength{\unitlength}{0.7cm}
  \begin{picture}(2,3)(0,0)
    \put (1,3){\line(0,-1){1.5}}
    \put (0,2){\line(1,0){2}}
    \put (0,2){\line(0,-1){2}}
    \put (1,2){\line(0,-1){2}}
    \put (2,2){\line(0,-1){2}}
    \put (0,2.2){\makebox(0,0){c}}
    \put (2,2.2){\makebox(0,0){i}}
  \end{picture}
  \hspace{.5cm}
  \setlength{\unitlength}{0.7cm}
  \begin{picture}(2,3)(0,0)
    \put (1,3){\line(0,-1){1}}
    \put (0,2){\line(1,0){2}}
    \put (0,2){\line(0,-1){1}}
    \put (1,2){\line(0,-1){1}}
    \put (2,2){\line(0,-1){2}}
    \put (0,1){\line(1,0){1}}
    \put (0.5,1){\line(0,-1){1}}
    \put (0,2.2){\makebox(0,0){c}}
    \put (2,2.2){\makebox(0,0){i}}
  \end{picture}
  \hspace{.5cm}
  \setlength{\unitlength}{0.7cm}
  \begin{picture}(2,3)(0,0)
    \put (1,3){\line(0,-1){1}}
    \put (0,2){\line(1,0){2}}
    \put (0,2){\line(0,-1){2}}
    \put (1,2){\line(0,-1){1}}
    \put (2,2){\line(0,-1){1}}
    \put (1,1){\line(1,0){1}}
    \put (1.5,1){\line(0,-1){1}}
    \put (0,2.2){\makebox(0,0){c}}
    \put (2,2.2){\makebox(0,0){i}}
  \end{picture}
  \hspace{.5cm}
  \setlength{\unitlength}{0.7cm}
  \begin{picture}(2,3)(0,0)
    \put (0,3){\line(0,-1){3}}
    \put (0,2){\line(1,0){2}}
    \put (1,2){\line(0,-1){2}}
    \put (2,2){\line(0,-1){1}}
    \put (2,1){\line(-1,0){1}}
    \put (1,2.2){\makebox(0,0){c}}
    \put (2,2.2){\makebox(0,0){i}}
  \end{picture}
  \caption{\label{fig3}All different possible events/genealogies in
    the ancestral selection graph (ASG) connected
    to~\eqref{eq:SDE}. Time in the ASG is running from top to bottom,
    i.e.\ from the present to the past. (A) In this case, when picking
    a single line in the present, its type only depends on the type of
    its ancestor, which is the single line at the bottom. (B) A
    T2-arrow leads to a split of a single line into a (straight)
    continuing line and an incoming line. In case~(i) (case~(ii)), if
    the incoming line has type $A$ (type~$B$), the type at the top is
    type $A$ (type~$B$) as well. If the incoming line has type~$B$
    (type~$A$), the type at the top equals the type of the continuing
    line. Here, the type of the line in the present depends on the
    types of both lines at the bottom. (C) A T3-arrow leads to a split
    of a single line in the checking (c), imitating~(i) and continuing
    line. Here, in case~(i) (case~(ii)), if the checking line is
    type~$A$ (in both cases), and the imitating branch is type~$B$
    (type~$A$), the type at the top is type~$B$ (type~$A$) as well. In
    all other cases, the type at the top equals the type of the
    continuing line. (D), (E), (F) In addition to a T3-branching
    event, coalescence of two lines might happen, leading to two lines
    present in the ASG at the bottom of the time window.}
\end{figure}
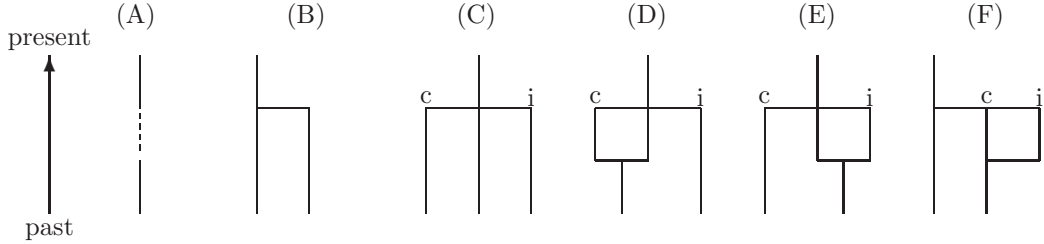

Once this graph is run for time~$t$ (i.e.\ from time $t$ down to time
0), and the starting frequency of type $A$ (at time~0) of the forward
process is given, we can determine the configuration of the $n$ lines.
First, assign randomly each of the lines of the ASG at time~0 with $A$
according to the starting frequency. Then, go through the ASG from
time 0 up to time~$t$. The types are inherited along every
T1-arrow. Moreover, we have to distinguish the inheritance rules in
the four cases, (i)--(iv).
\begin{enumerate}
\item[2A] In cases (i) and (iii), i.e.\ when $\beta>0$, the type of
  the incoming branch is inherited if it has type $A$. Otherwise the
  type of the continuing branch is inherited.
\item[2B] In cases (ii) and (iv), i.e.\ when $\beta<0$, the type of
  the incoming branch is inherited if it has type $B$.  Otherwise the
  type of the continuing branch is inherited.
\item[3A] In cases (i) and (iv), i.e.\ when $\gamma>0$, the type of
  the imitating branch is inherited if it has type $B$ and the
  checking branch has type $A$. Otherwise  the
  type of the continuing branch is inherited.
\item[3B] In cases (ii) and (iii), i.e.\ when $\gamma<0$, the type of
  the imitating branch is inherited if it has type $A$ and the
  checking branch has type $A$. Otherwise the type of the continuing
  branch is inherited.
\end{enumerate}
It is important to understand that these rules are reminiscent of the
corresponding transitions in the Moran model. As an example, consider
Figure~\ref{fig3}(D). (Here, the ASG is given starting with a single
line at the top, $n=1$.) In case (i) and (iv), the type of the line at
the top is $A$ only if both lines at the bottom have type $A$. If the
first line is $B$, the continuing branch of the T3-event is followed
which carries type $B$. If the first line is $A$ and the second line
is $B$, rule 3A says that the type of the imitating branch is
inherited, hence the line at the top has type $B$. In case (ii) and
(iii), the line at the top is $A$ if and only if the first line at the
bottom is type $A$. In this case, the imitating line can only be used
if it has type $A$ as well, leading to type $A$ at the top. If the
first line carries $B$, the continuing line is used at the T3-event,
leading to type $B$ at the top. In Table~\ref{tab1}, we assume that
the frequency of type $A$ at time 0 is $\varepsilon\ll 1$, and we give
all possibilities which lead to fixation of type $A$, which require at
most one type-$A$ individual in the past, and theire respective
probabilities up to order $\varepsilon$.

\section{Two proofs of Theorem~\ref{T:onethird}}
\label{S:twop}

\begin{table}
  \centering
  \begin{tabular}{|l||c|c|c|c|c|c|}\hline
    \vphantom{$\displaystyle\int\limits^0$}Genealogy $\mathcal G = g$
    & \setlength{\unitlength}{0,2cm}
    \begin{picture}(2,3)(0,0)
      \put (1,0){\line(0,1){1}}
      \put (1,2){\line(0,1){1}}
      \put (1,1.2){\line(0,1){.1}}
      \put (1,1.4){\line(0,1){.1}}
      \put (1,1.6){\line(0,1){.1}}
      \put (1,1.8){\line(0,1){.1}}
    \end{picture}
    &\setlength{\unitlength}{0,2cm}
    \begin{picture}(2,3)(0,0)
      \put (0,3){\line(0,-1){3}}
      \put (0,2){\line(1,0){1}}
      \put (1,2){\line(0,-1){2}}
    \end{picture}
    & \setlength{\unitlength}{0,2cm}
    \begin{picture}(2,3)(0,0)
      \put (1,3){\line(0,-1){1.5}}
      \put (0,2){\line(1,0){2}}
      \put (0,2){\line(0,-1){2}}
      \put (1,2){\line(0,-1){2}}
      \put (2,2){\line(0,-1){2}}
      \put (0,2.2){\makebox(0,0){\tiny c}}
      \put (2,2.2){\makebox(0,0){\tiny i}}
    \end{picture}
    &\setlength{\unitlength}{0,2cm}
    \begin{picture}(2,3)(0,0)
      \put (1,3){\line(0,-1){1}}
      \put (0,2){\line(1,0){2}}
      \put (0,2){\line(0,-1){1}}
      \put (1,2){\line(0,-1){1}}
      \put (2,2){\line(0,-1){2}}
      \put (0,1){\line(1,0){1}}
      \put (0.5,1){\line(0,-1){1}}
      \put (0,2.2){\makebox(0,0){\tiny c}}
      \put (2,2.2){\makebox(0,0){\tiny i}}
    \end{picture}
    &\setlength{\unitlength}{0,2cm}
    \begin{picture}(2,3)(0,0)
      \put (1,3){\line(0,-1){1}}
      \put (0,2){\line(1,0){2}}
      \put (0,2){\line(0,-1){2}}
      \put (1,2){\line(0,-1){1}}
      \put (2,2){\line(0,-1){1}}
      \put (1,1){\line(1,0){1}}
      \put (1.5,1){\line(0,-1){1}}
      \put (0,2.2){\makebox(0,0){\tiny c}}
      \put (2,2.2){\makebox(0,0){\tiny i}}
    \end{picture}
    &\setlength{\unitlength}{0,2cm}
    \begin{picture}(2,3)(0,0)
      \put (0,3){\line(0,-1){3}}
      \put (0,2){\line(1,0){2}}
      \put (1,2){\line(0,-1){2}}
      \put (2,2){\line(0,-1){1}}
      \put (2,1){\line(-1,0){1}}
      \put (1,2.2){\makebox(0,0){\tiny c}}
      \put (2,2.2){\makebox(0,0){\tiny i}}
    \end{picture}
    \\\hline\hline
    \vphantom{$\displaystyle\int\limits^0$}Case (i): in which cases does $A$ fix?
    & \setlength{\unitlength}{0,2cm}
    \begin{picture}(2,3)(0,0)
      \put (1,0){\line(0,1){1}}
      \put (1,2){\line(0,1){1}}
      \put (1,1.2){\line(0,1){.1}}
      \put (1,1.4){\line(0,1){.1}}
      \put (1,1.6){\line(0,1){.1}}
      \put (1,1.8){\line(0,1){.1}}
      \put (1,0){\makebox(0,0){\tiny$\bullet$}}
    \end{picture}
    &
    \setlength{\unitlength}{0,2cm}
    \begin{picture}(2,3)(0,0)
      \put (0,3){\line(0,-1){3}}
      \put (0,2){\line(1,0){1}}
      \put (1,2){\line(0,-1){2}}
      \put (0,0){\makebox(0,0){\tiny$\bullet$}}
    \end{picture}
    or
    \setlength{\unitlength}{0,2cm}
    \begin{picture}(2,3)(0,0)
      \put (0,3){\line(0,-1){3}}
      \put (0,2){\line(1,0){1}}
      \put (1,2){\line(0,-1){2}}
      \put (1,0){\makebox(0,0){\tiny$\bullet$}}
    \end{picture}
    & \setlength{\unitlength}{0,2cm}
    \begin{picture}(2,3)(0,0)
      \put (1,3){\line(0,-1){1.5}}
      \put (0,2){\line(1,0){2}}
      \put (0,2){\line(0,-1){2}}
      \put (1,2){\line(0,-1){2}}
      \put (2,2){\line(0,-1){2}}
      \put (0,2.2){\makebox(0,0){\tiny c}}
      \put (2,2.2){\makebox(0,0){\tiny i}}
      \put (1,0){\makebox(0,0){\tiny$\bullet$}}
    \end{picture}
    &\setlength{\unitlength}{0,2cm}
    \begin{picture}(2,3)(0,0)
      \put (1,3){\line(0,-1){1}}
      \put (0,2){\line(1,0){2}}
      \put (0,2){\line(0,-1){1}}
      \put (1,2){\line(0,-1){1}}
      \put (2,2){\line(0,-1){2}}
      \put (0,1){\line(1,0){1}}
      \put (0.5,1){\line(0,-1){1}}
      \put (0,2.2){\makebox(0,0){\tiny c}}
      \put (2,2.2){\makebox(0,0){\tiny i}}
    \end{picture}
    &\setlength{\unitlength}{0,2cm}
    \begin{picture}(2,3)(0,0)
      \put (1,3){\line(0,-1){1}}
      \put (0,2){\line(1,0){2}}
      \put (0,2){\line(0,-1){2}}
      \put (1,2){\line(0,-1){1}}
      \put (2,2){\line(0,-1){1}}
      \put (1,1){\line(1,0){1}}
      \put (1.5,1){\line(0,-1){1}}
      \put (0,2.2){\makebox(0,0){\tiny c}}
      \put (2,2.2){\makebox(0,0){\tiny i}}
      \put (1.5,0){\makebox(0,0){\tiny$\bullet$}}
    \end{picture}
    &\setlength{\unitlength}{0,2cm}
    \begin{picture}(2,3)(0,0)
      \put (0,3){\line(0,-1){3}}
      \put (0,2){\line(1,0){2}}
      \put (1,2){\line(0,-1){2}}
      \put (2,2){\line(0,-1){1}}
      \put (2,1){\line(-1,0){1}}
      \put (1,2.2){\makebox(0,0){\tiny c}}
      \put (2,2.2){\makebox(0,0){\tiny i}}
      \put (0,0){\makebox(0,0){\tiny$\bullet$}}
    \end{picture}
    \\\hline
    \vphantom{$\displaystyle\int$}Case (i): $\mathbf P[A\text{ fixes}|\; \mathcal G = g, X_0=\varepsilon]$&
    $\varepsilon$&
    $2\varepsilon$&
    $\varepsilon$&
    $0$&
    $\varepsilon$&
    $\varepsilon$
    \\\hline\hline
    \vphantom{$\displaystyle\int\limits^0$}Case (ii): in which cases does $A$ fix?
    & \setlength{\unitlength}{0,2cm}
    \begin{picture}(2,3)(0,0)
      \put (1,0){\line(0,1){1}}
      \put (1,2){\line(0,1){1}}
      \put (1,1.2){\line(0,1){.1}}
      \put (1,1.4){\line(0,1){.1}}
      \put (1,1.6){\line(0,1){.1}}
      \put (1,1.8){\line(0,1){.1}}
      \put (1,0){\makebox(0,0){\tiny$\bullet$}}
    \end{picture}
    &\setlength{\unitlength}{0,2cm}
    \begin{picture}(2,3)(0,0)
      \put (0,3){\line(0,-1){3}}
      \put (0,2){\line(1,0){1}}
      \put (1,2){\line(0,-1){2}}
    \end{picture}
    & \setlength{\unitlength}{0,2cm}
    \begin{picture}(3,3)(0,0)
      \put (1,3){\line(0,-1){1.5}}
      \put (0,2){\line(1,0){2}}
      \put (0,2){\line(0,-1){2}}
      \put (1,2){\line(0,-1){2}}
      \put (2,2){\line(0,-1){2}}
      \put (0,2.2){\makebox(0,0){\tiny c}}
      \put (2,2.2){\makebox(0,0){\tiny i}}
      \put (1,0){\makebox(0,0){\tiny$\bullet$}}
    \end{picture}
    &\setlength{\unitlength}{0,2cm}
    \begin{picture}(2,3)(0,0)
      \put (1,3){\line(0,-1){1}}
      \put (0,2){\line(1,0){2}}
      \put (0,2){\line(0,-1){1}}
      \put (1,2){\line(0,-1){1}}
      \put (2,2){\line(0,-1){2}}
      \put (0,1){\line(1,0){1}}
      \put (0.5,1){\line(0,-1){1}}
      \put (0,2.2){\makebox(0,0){\tiny c}}
      \put (2,2.2){\makebox(0,0){\tiny i}}
      \put (.5,0){\makebox(0,0){\tiny$\bullet$}}
    \end{picture}
    &\setlength{\unitlength}{0,2cm}
    \begin{picture}(2,3)(0,0)
      \put (1,3){\line(0,-1){1}}
      \put (0,2){\line(1,0){2}}
      \put (0,2){\line(0,-1){2}}
      \put (1,2){\line(0,-1){1}}
      \put (2,2){\line(0,-1){1}}
      \put (1,1){\line(1,0){1}}
      \put (1.5,1){\line(0,-1){1}}
      \put (0,2.2){\makebox(0,0){\tiny c}}
      \put (2,2.2){\makebox(0,0){\tiny i}}
      \put (1.5,0){\makebox(0,0){\tiny$\bullet$}}
    \end{picture}
    &\setlength{\unitlength}{0,2cm}
    \begin{picture}(2,3)(0,0)
      \put (0,3){\line(0,-1){3}}
      \put (0,2){\line(1,0){2}}
      \put (1,2){\line(0,-1){2}}
      \put (2,2){\line(0,-1){1}}
      \put (2,1){\line(-1,0){1}}
      \put (1,2.2){\makebox(0,0){\tiny c}}
      \put (2,2.2){\makebox(0,0){\tiny i}}
      \put (0,0){\makebox(0,0){\tiny$\bullet$}}
    \end{picture}
    or
    \setlength{\unitlength}{0,2cm}
    \begin{picture}(2,3)(0,0)
      \put (0,3){\line(0,-1){3}}
      \put (0,2){\line(1,0){2}}
      \put (1,2){\line(0,-1){2}}
      \put (2,2){\line(0,-1){1}}
      \put (2,1){\line(-1,0){1}}
      \put (1,2.2){\makebox(0,0){\tiny c}}
      \put (2,2.2){\makebox(0,0){\tiny i}}
      \put (1,0){\makebox(0,0){\tiny$\bullet$}}
    \end{picture}
    \\\hline
    \vphantom{$\displaystyle\int$}Case (ii): $\mathbf P[A\text{ fixes}| \;\mathcal G = g, X_0=\varepsilon]$&
    $\varepsilon$&
    0&
    $\varepsilon$&
    $\varepsilon$&
    $\varepsilon$&
    $2\varepsilon$
    \\\hline\hline
    \vphantom{$\displaystyle\int\limits^0$}Case (iii): in which cases does $A$ fix?
    & \setlength{\unitlength}{0,2cm}
    \begin{picture}(2,3)(0,0)
      \put (1,0){\line(0,1){1}}
      \put (1,2){\line(0,1){1}}
      \put (1,1.2){\line(0,1){.1}}
      \put (1,1.4){\line(0,1){.1}}
      \put (1,1.6){\line(0,1){.1}}
      \put (1,1.8){\line(0,1){.1}}
      \put (1,0){\makebox(0,0){\tiny$\bullet$}}
    \end{picture}
    &
    \setlength{\unitlength}{0,2cm}
    \begin{picture}(2,3)(0,0)
      \put (0,3){\line(0,-1){3}}
      \put (0,2){\line(1,0){1}}
      \put (1,2){\line(0,-1){2}}
      \put (0,0){\makebox(0,0){\tiny$\bullet$}}
    \end{picture}
    or
    \setlength{\unitlength}{0,2cm}
    \begin{picture}(2,3)(0,0)
      \put (0,3){\line(0,-1){3}}
      \put (0,2){\line(1,0){1}}
      \put (1,2){\line(0,-1){2}}
      \put (1,0){\makebox(0,0){\tiny$\bullet$}}
    \end{picture}
    & \setlength{\unitlength}{0,2cm}
    \begin{picture}(3,3)(0,0)
      \put (1,3){\line(0,-1){1.5}}
      \put (0,2){\line(1,0){2}}
      \put (0,2){\line(0,-1){2}}
      \put (1,2){\line(0,-1){2}}
      \put (2,2){\line(0,-1){2}}
      \put (0,2.2){\makebox(0,0){\tiny c}}
      \put (2,2.2){\makebox(0,0){\tiny i}}
      \put (1,0){\makebox(0,0){\tiny$\bullet$}}
    \end{picture}
    &\setlength{\unitlength}{0,2cm}
    \begin{picture}(2,3)(0,0)
      \put (1,3){\line(0,-1){1}}
      \put (0,2){\line(1,0){2}}
      \put (0,2){\line(0,-1){1}}
      \put (1,2){\line(0,-1){1}}
      \put (2,2){\line(0,-1){2}}
      \put (0,1){\line(1,0){1}}
      \put (0.5,1){\line(0,-1){1}}
      \put (0,2.2){\makebox(0,0){\tiny c}}
      \put (2,2.2){\makebox(0,0){\tiny i}}
      \put (.5,0){\makebox(0,0){\tiny$\bullet$}}
    \end{picture}
    &\setlength{\unitlength}{0,2cm}
    \begin{picture}(2,3)(0,0)
      \put (1,3){\line(0,-1){1}}
      \put (0,2){\line(1,0){2}}
      \put (0,2){\line(0,-1){2}}
      \put (1,2){\line(0,-1){1}}
      \put (2,2){\line(0,-1){1}}
      \put (1,1){\line(1,0){1}}
      \put (1.5,1){\line(0,-1){1}}
      \put (0,2.2){\makebox(0,0){\tiny c}}
      \put (2,2.2){\makebox(0,0){\tiny i}}
      \put (1.5,0){\makebox(0,0){\tiny$\bullet$}}
    \end{picture}
    &\setlength{\unitlength}{0,2cm}
    \begin{picture}(2,3)(0,0)
      \put (0,3){\line(0,-1){3}}
      \put (0,2){\line(1,0){2}}
      \put (1,2){\line(0,-1){2}}
      \put (2,2){\line(0,-1){1}}
      \put (2,1){\line(-1,0){1}}
      \put (1,2.2){\makebox(0,0){\tiny c}}
      \put (2,2.2){\makebox(0,0){\tiny i}}
      \put (0,0){\makebox(0,0){\tiny$\bullet$}}
    \end{picture}
    or
    \setlength{\unitlength}{0,2cm}
    \begin{picture}(2,3)(0,0)
      \put (0,3){\line(0,-1){3}}
      \put (0,2){\line(1,0){2}}
      \put (1,2){\line(0,-1){2}}
      \put (2,2){\line(0,-1){1}}
      \put (2,1){\line(-1,0){1}}
      \put (1,2.2){\makebox(0,0){\tiny c}}
      \put (2,2.2){\makebox(0,0){\tiny i}}
      \put (1,0){\makebox(0,0){\tiny$\bullet$}}
    \end{picture}
    \\\hline
    \vphantom{$\displaystyle\int$}Case (iii): $\mathbf P[A\text{ fixes}| \;\mathcal G = g, X_0=\varepsilon]$&
    $\varepsilon$&
    $2\varepsilon$&
    $\varepsilon$&
    $\varepsilon$&
    $\varepsilon$&
    $2\varepsilon$
    \\\hline\hline
    \vphantom{$\displaystyle\int\limits^0$}Case (iv): in which cases does $A$ fix?
    & \setlength{\unitlength}{0,2cm}
    \begin{picture}(2,3)(0,0)
      \put (1,0){\line(0,1){1}}
      \put (1,2){\line(0,1){1}}
      \put (1,1.2){\line(0,1){.1}}
      \put (1,1.4){\line(0,1){.1}}
      \put (1,1.6){\line(0,1){.1}}
      \put (1,1.8){\line(0,1){.1}}
      \put (1,0){\makebox(0,0){\tiny$\bullet$}}
    \end{picture}
    &\setlength{\unitlength}{0,2cm}
    \begin{picture}(2,3)(0,0)
      \put (0,3){\line(0,-1){3}}
      \put (0,2){\line(1,0){1}}
      \put (1,2){\line(0,-1){2}}
    \end{picture}
    & \setlength{\unitlength}{0,2cm}
    \begin{picture}(2,3)(0,0)
      \put (1,3){\line(0,-1){1.5}}
      \put (0,2){\line(1,0){2}}
      \put (0,2){\line(0,-1){2}}
      \put (1,2){\line(0,-1){2}}
      \put (2,2){\line(0,-1){2}}
      \put (0,2.2){\makebox(0,0){\tiny c}}
      \put (2,2.2){\makebox(0,0){\tiny i}}
      \put (1,0){\makebox(0,0){\tiny$\bullet$}}
    \end{picture}
    &\setlength{\unitlength}{0,2cm}
    \begin{picture}(2,3)(0,0)
      \put (1,3){\line(0,-1){1}}
      \put (0,2){\line(1,0){2}}
      \put (0,2){\line(0,-1){1}}
      \put (1,2){\line(0,-1){1}}
      \put (2,2){\line(0,-1){2}}
      \put (0,1){\line(1,0){1}}
      \put (0.5,1){\line(0,-1){1}}
      \put (0,2.2){\makebox(0,0){\tiny c}}
      \put (2,2.2){\makebox(0,0){\tiny i}}
    \end{picture}
    &\setlength{\unitlength}{0,2cm}
    \begin{picture}(2,3)(0,0)
      \put (1,3){\line(0,-1){1}}
      \put (0,2){\line(1,0){2}}
      \put (0,2){\line(0,-1){2}}
      \put (1,2){\line(0,-1){1}}
      \put (2,2){\line(0,-1){1}}
      \put (1,1){\line(1,0){1}}
      \put (1.5,1){\line(0,-1){1}}
      \put (0,2.2){\makebox(0,0){\tiny c}}
      \put (2,2.2){\makebox(0,0){\tiny i}}
      \put (1.5,0){\makebox(0,0){\tiny$\bullet$}}
    \end{picture}
    &\setlength{\unitlength}{0,2cm}
    \begin{picture}(2,3)(0,0)
      \put (0,3){\line(0,-1){3}}
      \put (0,2){\line(1,0){2}}
      \put (1,2){\line(0,-1){2}}
      \put (2,2){\line(0,-1){1}}
      \put (2,1){\line(-1,0){1}}
      \put (1,2.2){\makebox(0,0){\tiny c}}
      \put (2,2.2){\makebox(0,0){\tiny i}}
      \put (0,0){\makebox(0,0){\tiny$\bullet$}}
    \end{picture}
    \\\hline
    \vphantom{$\displaystyle\int$}Case (iv): $\mathbf P[A\text{ fixes}| \;\mathcal G = g, X_0=\varepsilon]$&
    $\varepsilon$&
    0&
    $\varepsilon$&
    $0$&
    $\varepsilon$&
    $\varepsilon$
    \\\hline
  \end{tabular}
  \caption{\label{tab1}All possible genealogies close to time $0$ in the ASG for small 
    selection strength $\alpha$ and
    a small starting frequency $X_0=\varepsilon\ll 1$. In the second (fourth, sixth, eighth) line we indicate
    which of the lines at the bottom has to be type $A$ such that type $A$ fixes in case (i) (case (ii), 
    case (iii), case (iv)).
    In the third (fifth, seventh, ninth) line we give the corresponding probabilities 
    (with error terms of order $\varepsilon^2$).}
\end{table}

\subsection{A proof based on \cite{Rousset2003} and \cite{Ladret2007}}
Although the proof of \cite{Ladret2007}, which is based on ideas from
\cite{Rousset2003} is carried out in a time-discrete model, the same
argument works for the SDE~\eqref{eq:SDE}. Again, we write $\mathbf
E_{\varepsilon,\alpha}[.]$ for expectations with $X_0=\varepsilon$ and
fitness coefficient $\alpha$. First, \eqref{eq:SDE} is continuous in
the parameter $\alpha$, such that $\mathbf
E_{\varepsilon,\alpha}[f(X_t)] = \mathbf
E_{\varepsilon,0}[f(X_t)](1+\mathcal O(\alpha))$ for $f\in\mathcal
C_b^2(\mathbb R)$. Moreover, by the duality of the Wright--Fisher
diffusion, i.e.\ \eqref{eq:SDE} with $\alpha=0$, to Kingman's
coalescent,
\begin{align*}
  \mathbf E_{\varepsilon,0}[X_t(1-X_t)] & = e^{-t}
  \varepsilon(1-\varepsilon) = e^{-t} \big(\varepsilon +
  \mathcal O(\varepsilon^2)\big),\\
  \mathbf E_{\varepsilon,0}[X_t^2(1-X_t)] & =
  e^{-3t}\varepsilon^2(1-\varepsilon) + \int_0^t 3e^{-3s} e^{-(t-s)}
  \frac 13 \varepsilon(1-\varepsilon) ds \\ & = \frac 12
  e^{-t}(1-e^{-2t})\big(\varepsilon + \mathcal O(\varepsilon^2)\big)
\end{align*}
as $\varepsilon\to 0$. Here, the first equation arises since
$2X_t(1-X_t)$ is the probability to obtain two different types when
picking from the population. This event requires that the two sampled
individuals do not share an ancestor between times~0 and
$t$. Moreover, both ancestors have to have different types. For the
second equality, when sampling three individuals, two of which are
type~$A$, there are two possibilities. Either no coalescence of the
three sampled lines occurred between times~0 and $t$, or two of three
lines coalesced and have a type~$A$-ancestor. Combining these results,
\begin{align*}
  \mathbf P_{\varepsilon,\alpha}(A \text{ fixes}) & = \mathbf
  E_{\varepsilon, \alpha}[X_\infty] = \varepsilon + \int_0^\infty
  \frac{d}{dt} \mathbf E_{\varepsilon, \alpha}[X_t]dt \\ & =
  \varepsilon + \alpha \int_0^\infty \mathbf
  E_{\varepsilon,0}[X_t(1-X_t)(\beta - \gamma X_t)](1+\mathcal
  O(\alpha)) dt \\ & = \varepsilon + \alpha \int_0^\infty \big(\beta
  e^{-t}\big(\varepsilon + \mathcal O(\varepsilon^2)\big) - \frac 12
  \gamma e^{-t}(1-e^{-2t})\big(\varepsilon + \mathcal
  O(\varepsilon^2)\big)\big)(1+\mathcal O(\alpha))dt \\ & =
  \varepsilon\Big( 1 + \alpha \Big(\beta - \gamma \frac 13\Big) \Big)
  + \mathcal O(\alpha^2, \varepsilon^2),
\end{align*}
which finishes the first proof.

\subsection{A proof based on the ASG}
Let us prove Theorem~\ref{T:onethird} by using the ASG.  Here, type
$A$ eventually fixes if and only if the common ancestor of all
individuals carries type $A$. Therefore, we have to study the ASG
which has run for an infinite amount of time (i.e.\ from time infinity
back to time~0). Since $\alpha$ is assumed to be small, we can assume
that the ASG has only a single line at time~0 with high
probability. However, with probability of order $\alpha$, the ASG has
split close to time~0 but the resulting lines have not fully coalesced
yet. Since $\alpha$ is small, we can assume that the last split event
happened when the ASG only had a single line.

In Figure~\ref{fig3}, we list all possible ASGs near time~0 which we
have to consider. Start with case (B), which is a split in two
lines. We abbreviate this genealogy by $\zwei$. Noting that such split
events occur at rate $\alpha|\beta|$, and coalescence of the resulting
two lines happens at rate~1, the probability for this case is (by
competing exponential clocks)
\begin{align}\label{eq:zwei}
  \mathbf P\Big( \zwei \Big) & =
  \frac{\alpha|\beta|}{1+\alpha|\beta|}+ \mathcal O(\alpha^2) =
  \alpha|\beta| + \mathcal O(\alpha^2).
\end{align}
Similarly, consider the case when the last split occurs into three
lines, which do not coalesce up to time~0 (case $\dreinocoal$). Since
coalescence of any pair happens at rate~3, we obtain
\begin{align}\label{eq:dreinocoal}
  \mathbf P\Big(\dreinocoal\Big) & =
  \frac{\alpha|\gamma|}{3+\alpha|\gamma|} + \mathcal O(\alpha^2)=
  \frac{\alpha|\gamma|}{3} + \mathcal O(\alpha^2).
\end{align}
Finally, the last split can lead to three lines and two of them
coalesce up to time~0. Since all three possible genealogies (denoted
$\dreionecoala, \dreionecoalb$ and $\dreionecoalc$) have equal
probability, we obtain
\begin{align}\label{eq:dreionecoal}
  \mathbf P\Big( \dreionecoala\Big) & = \mathbf P\Big(
  \dreionecoalb\Big) = \mathbf P\Big( \dreionecoalc \Big) = \frac 13
  \frac{\alpha|\gamma|}{1+\alpha|\gamma|} + \mathcal O(\alpha^2) =
  \frac{\alpha|\gamma|}{3} + \mathcal O(\alpha^2).
\end{align}
Last, we have for the remaining case $\eins$
\begin{equation}\label{eq:eins}
  \begin{aligned}
    \mathbf P\Big( \eins \Big) & = 1 - \mathbf P\Big( \zwei \Big) -
    \mathbf P\Big(\dreinocoal \Big) - \mathbf P\Big(
    \dreionecoala\Big) - \mathbf P\Big( \dreionecoalb\Big) - \mathbf
    P\Big( \dreionecoalc \Big) + \mathcal O(\alpha^2)\\ & = 1 - \alpha
    \Big(|\beta| + \frac 43 |\gamma|\Big) + \mathcal O(\alpha^2).
  \end{aligned}
\end{equation}
Using Table~\ref{tab1} for the probabilities of fixation conditioned
on the genealogy, we can now collect all terms in the four cases. We
obtain
\begin{align*}
  \text{Case (i): } & p_\text{fix}(\varepsilon,\alpha) = \varepsilon
  \cdot \mathbf P\Big( \eins \Big) + 2\varepsilon \cdot \mathbf P\Big(
  \zwei \Big) + \varepsilon\cdot \mathbf P\Big( \dreinocoal \Big) +
  \varepsilon \cdot \mathbf P\Big( \dreionecoalb \Big) \\ & \qquad
  \qquad \qquad \qquad \qquad \qquad \qquad \qquad \qquad \qquad +
  \varepsilon\cdot \mathbf P\Big(\dreionecoalc \Big) + \mathcal
  O(\alpha^2,
  \varepsilon^2)\\
  \text{Case (ii): } & p_\text{fix}(\varepsilon,\alpha) = \varepsilon
  \cdot \mathbf P\Big( \eins \Big) + \varepsilon\cdot \mathbf P\Big(
  \dreinocoal\Big) + \varepsilon \cdot \mathbf P\Big(
  \dreionecoala\Big) + \varepsilon \cdot \mathbf P\Big(\dreionecoalb
  \Big) \\ & \qquad \qquad \qquad \qquad \qquad \qquad \qquad \qquad
  \qquad \qquad + 2\varepsilon\cdot \mathbf P\Big(\dreionecoalc \Big)
  + \mathcal O(\alpha^2, \varepsilon^2) \\ \text{Case (iii): } &
  p_\text{fix}(\varepsilon,\alpha) = \varepsilon \cdot \mathbf P\Big(
  \eins \Big) + 2\varepsilon \cdot \mathbf P\Big( \zwei \Big) +
  \varepsilon\cdot \mathbf P\Big( \dreinocoal \Big) + \varepsilon
  \cdot \mathbf P\Big( \dreionecoala \Big) \\ & \qquad \qquad \qquad
  \qquad \qquad \qquad \qquad \qquad + \varepsilon \cdot \mathbf
  P\Big( \dreionecoalb \Big) + 2\varepsilon\cdot \mathbf
  P\Big(\dreionecoalc \Big) + \mathcal O(\alpha^2, \varepsilon^2) \\
  \text{Case (iv): } & p_\text{fix}(\varepsilon,\alpha) = \varepsilon
  \cdot \mathbf P\Big( \eins \Big) + \varepsilon\cdot \mathbf P\Big(
  \dreinocoal \Big) + \varepsilon \cdot \mathbf P\Big( \dreionecoalb
  \Big) + \varepsilon\cdot \mathbf P\Big(\dreionecoalc \Big) +
  \mathcal O(\alpha^2, \varepsilon^2)
\end{align*}
Plugging in the probabilities from~\eqref{eq:zwei},
\eqref{eq:dreinocoal}, \eqref{eq:dreionecoal} and \eqref{eq:eins}, we
obtain in all cases
\begin{equation}
  \begin{aligned}
    p_\text{fix}(\varepsilon,\alpha) = \varepsilon\Big( 1 + \alpha
    \Big( \beta - \gamma \frac 13 \Big)\Big) + \mathcal O(\alpha^2,
    \varepsilon^2)
  \end{aligned}
\end{equation}
which finishes the second proof of Theorem~\ref{T:onethird}.

\section{Proof of Theorem~\ref{T:dist}}
\label{S:proof2}
The proof is based on the graphical construction of the Moran model;
see Figure~\ref{fig2}. We have to take into account all three
different kinds of arrows. Let $R_{12} = R_{12}(t)$ be the random
distance of two individuals taken at random from the Moran model at
time $t$. In addition, $Z_3 = Z_3(t)$ and $Z_4 = Z_4(t)$ are the types
of two additionally sampled individuals. We claim that in case (i)
\begin{equation}
  \label{eq:distbasic}
  \begin{aligned}
    \frac{d}{dt} \mathbf E_{N,\alpha}[e^{-\lambda R_{12}}] & =
    -\lambda \mathbf E_{N,\alpha}[e^{-\lambda R_{12}}] + 1 - \mathbf
    E_{N,\alpha}[e^{-\lambda R_{12}}]  \\
    & \qquad + \alpha |\beta| \cdot \mathbf
    E_{N,\alpha}[1_{Z_3=A} (e^{-\lambda R_{23}} - e^{-\lambda R_{12}})] \\
    & \qquad \qquad \qquad + \alpha|\gamma| \cdot \mathbf E_{N,\alpha}
    [1_{Z_3=B}1_{Z_4=A} (e^{-\lambda R_{23}} - e^{-\lambda R_{12}})] +
    \mathcal O(1/N).
  \end{aligned}
\end{equation}
Here, the first term describes the increase of the genealogical
distance of individuals~1 and~2 if no events occur. Second, T1-arrows
between~1 and~2 lead to $R_{12}=0$ (hence $e^{-\lambda R_{12}}=1$) at
rate~1. Third, the origin of T2-arrows is with high probability
($1-\mathcal O(1/N)$) neither individual~1 nor~2, but a third
individual, with type $Z_3$. However, this arrow only takes effect if
$Z_3=A$. Last, T3-arrows again originate from a third individual with
type $Z_3$. It picks a fourth individual with type $Z_4$ and if
$Z_3=B$ and $Z_4=A$, the arrow takes effect. All terms require that
the two/three/four chosen individuals are distinct; hence the error
term $\mathcal O(1/N)$.

Our goal is to approximate the right hand side
of~\eqref{eq:distbasic}. We note that the ASG implies that for all
expectations on the right hand side we have that $\mathbf
E_{N,\alpha}[.] = \mathbf E_{N,0}[.] + \mathcal O(\alpha)$ for small
$\alpha$ and large $N$. Hence, we can use $\mathbf E_{N,0}[.]$ as an
approximation which is valid up to order $\alpha$.

Recall that the mutation rate from $A$ to $B$ is $\theta_B/2$ and from
$B$ to $A$ is $\theta_A/2$, as well as $\bar\theta := \theta_A +
\theta_B$. We note that for $N\to\infty$, we can use Kingman's
coalescent, since it is known to give genealogies in the large
population limit of the neutral Moran model. We write $\mathbf E[.]$
for the expectation when using Kingman's coalescent and finally argue
that $\mathbf E_{N,0}[.] = \mathbf E[.] + \mathcal O(1/N)$. We obtain
\begin{align*}
  \mathbf E[1_{Z_1=B}1_{Z_2=A}] & = \frac{\theta_B/2}{1+\bar\theta}
  \cdot \frac{\theta_A}{\bar\theta} + \frac{\theta_A/2}{1+\bar\theta}
  \cdot \frac{\theta_B}{\bar\theta} =
  \frac{\theta_A\theta_B}{\bar\theta}\frac{1}{1+\bar\theta},\\
  \mathbf E[1_{Z_1=A} e^{-\lambda R_{12}}] & = \mathbf
  E[1_{Z_3=A}e^{-\lambda R_{12}}] = \frac{\theta_A}{\bar\theta}
  \frac{1}{1+\lambda},\\
  \mathbf E[1_{Z_1=B} e^{-\lambda R_{12}}] & = \mathbf
  E[1_{Z_3=B}e^{-\lambda R_{12}}] = \frac{\theta_B}{\bar\theta}
  \frac{1}{1+\lambda}.
\end{align*}
Here, the first equality holds since $Z_1=B$ and $Z_2=A$ if and only
if the first mutation event occurs before the two sampled lines
coalesce. This mutation event must either be $B\to A$ in individual 1
or $B\to A$ in individual 2. Using similar arguments, we obtain
\begin{align*}
  \mathbf E[1_{Z_1=A} 1_{Z_2=B} e^{-\lambda R_{12}}] & = \int_0^\infty
  (1+\bar\theta)e^{-(1+\bar\theta)t}
  \cdot\Big(\frac{1 + \theta_A/2 + \theta_B/2}{1+\bar\theta}\cdot 0 \\
  & \qquad\qquad+ \frac{\theta_A/2}{1+\bar\theta} \mathbf
  E[1_{Z_2=B}e^{-\lambda(t+R_{12})}] +
  \frac{\theta_B/2}{1+\bar\theta}\mathbf
  E[1_{Z_1=A}e^{-\lambda(t+R_{12})}]\Big)dt \\ & =
  \frac{\theta_A\theta_B}{\bar\theta}\frac{1}{(1+\bar\theta
    + \lambda)(1+\lambda)},\\
  \mathbf E[1_{Z_2=B} 1_{Z_3=A} e^{-\lambda R_{12}}] & = \int_0^\infty
  (3+\bar\theta)e^{-(3+\bar\theta)t} \\ & \qquad \cdot\Big(
  \frac{1}{3+\bar\theta} \mathbf E[1_{Z_1=A} 1_{Z_2=B} e^{-\lambda
    (t+R_{12})}] + \frac{1}{3+\bar\theta} \mathbf E[1_{Z_2=A}
  1_{Z_3=B} e^{-\lambda t}] \\ & \qquad \qquad +
  \frac{\theta_A/2}{3+\bar\theta}\mathbf E[1_{Z_1=B}e^{-\lambda
    (t+R_{12})}] + \frac{\theta_B/2}{3+\bar\theta}\mathbf
  E[1_{Z_3=A}e^{-\lambda (t+R_{12})}]\Big) dt \\ & =
  \frac{\theta_A\theta_B}{\bar\theta}\frac{1}{3+\bar\theta+\lambda}\Big(
  \frac{1}{(1+\bar\theta+\lambda)(1+\lambda)} + \frac{1}{1+\bar\theta}
  + \frac{1}{1+\lambda}\Big) \\ & =
  \frac{\theta_A\theta_B}{\bar\theta}\Big(
  \frac{2+\bar\theta+\lambda}{3+\bar\theta+\lambda}\frac{1}{(1+\bar\theta+\lambda)(1+\lambda)}
  + \frac{1}{(3+\bar\theta+\lambda)(\bar\theta+1)}\Big),\\
  \mathbf E[1_{Z_2=A} 1_{Z_3=B} e^{-\lambda R_{12}}] & =
  \mathbf E[1_{Z_2=B} 1_{Z_3=A} e^{-\lambda R_{12}}],
\end{align*}
\begin{align*}
  \mathbf E[1_{Z_3=B} 1_{Z_4=A} e^{-\lambda R_{12}}] & = \int_0^\infty
  (6+\bar\theta)e^{-(6+\bar\theta)t} \\ & \qquad \cdot\Big(
  \frac{1}{6+\bar\theta} \mathbf E[1_{Z_3=B} 1_{Z_4=A}e^{-\lambda t}]
  + \frac{4}{6+\bar\theta} \mathbf E[1_{Z_2=A} 1_{Z_3=B} e^{-\lambda
    (t+R_{12})}] \\ & \qquad \qquad + \frac{\theta_A/2}{6+\bar\theta}
  \mathbf E[1_{Z_3=B} e^{-\lambda (t+R_{12})}] +
  \frac{\theta_B/2}{6+\bar\theta} \mathbf E[1_{Z_3=A} e^{-\lambda
    (t+R_{12})}]\Big) dt \\ & = \frac{\theta_A\theta_B}{\bar\theta}
  \frac{1}{6+\bar\theta + \lambda} \Big(\frac{1}{1+\bar\theta} +
  \frac{2+\bar\theta+\lambda}{3+\bar\theta+\lambda}\frac{4}{(1+\bar\theta+\lambda)(1+\lambda)}
  \\ & \qquad \qquad \qquad \qquad \qquad \qquad \qquad \qquad +
  \frac{4}{(3+\bar\theta + \lambda)(1+\bar\theta)}+
  \frac{1}{1+\lambda} \Big) \\ & = \frac{\theta_A\theta_B}{\bar\theta}
  \Big(\frac{7+\bar\theta+\lambda}{6 + \bar\theta +
    \lambda}\frac{1}{(3+\bar\theta+\lambda)(1+\bar\theta)} \\ & \qquad
  \qquad \qquad + \frac{2+\bar\theta+\lambda}{3+\bar\theta+\lambda}
  \frac{5 + \bar\theta + \lambda}{6 + \bar\theta +
    \lambda}\frac{1}{(1+\bar\theta+\lambda)(1+\lambda)} \\ & \qquad
  \qquad \qquad \qquad \qquad \qquad +
  \frac{1}{(6+\bar\theta+\lambda)(3+\bar\theta+\lambda)(1+\lambda)}\Big)
\end{align*}
Plugging the last two terms in \eqref{eq:distbasic} we find that in
equilibrium
\begin{align*}
  \mathbf E_{N,\alpha} [e^{-\lambda R_{12}}] & = \frac{1}{1+\lambda} +
  \frac{\alpha|\gamma|}{1+\lambda} \mathbf
  E_{N,0}[1_{Z_2=A}1_{Z_3=B}e^{-\lambda
    R_{12}}-1_{Z_3=A}1_{Z_4=B}e^{-\lambda R_{12}}] + \mathcal
  O(\alpha^2) \\ & = \frac{1}{1+\lambda} +
  \frac{\alpha|\gamma|}{1+\lambda}\frac{\theta_A\theta_B}{\bar \theta}
  \frac{1}{6+\bar\theta + \lambda} \Big( \frac{2 + \bar\theta +
    \lambda}{3+\bar\theta+\lambda}
  \frac{1}{(1+\bar\theta+\lambda)(1+\lambda)} \\ & \qquad \qquad
  \qquad \qquad \qquad \qquad -
  \frac{1}{(3+\bar\theta+\lambda)(1+\bar\theta)} -
  \frac{1}{(3+\bar\theta+\lambda)(1 + \lambda)}\Big) + \mathcal
  O(\alpha^2) \\ & = \frac{1}{1+\lambda} +
  \frac{\alpha|\gamma|}{1+\lambda}\frac{\theta_A\theta_B}{\bar \theta}
  \frac{1}{6+\bar\theta + \lambda}
  \frac{2+\bar\theta+\lambda}{3+\bar\theta+\lambda} \Big(
  \frac{1}{(1+\bar\theta+\lambda)(1+\lambda)} \\ & \qquad \qquad
  \qquad \qquad \qquad \qquad \qquad \qquad \qquad \qquad -
  \frac{1}{(1+\bar\theta)(1+\lambda)}\Big) + \mathcal O(\alpha^2) \\ &
  = \frac{1}{1+\lambda} - \alpha|\gamma|\frac{\theta_A\theta_B}{\bar
    \theta} \frac{(2+\bar\theta+\lambda)\lambda}{(6+\bar\theta +
    \lambda)(3+\bar\theta+\lambda)(1+\bar\theta +
    \lambda)(1+\bar\theta)(1+\lambda)^2} + \mathcal O(\alpha^2).
\end{align*}
In particular, using a derivative according to $\lambda$ at
$\lambda=0$,
\begin{align*}
  \mathbf E_{N,\alpha}[R_{12}] & = - \frac{\partial}{\partial\lambda}
  \mathbf E_{N,\alpha} [e^{-\lambda R_{12}}]\Big|_{\lambda=0} = 1 +
  \alpha|\gamma|\frac{\theta_A\theta_B}{\bar\theta}
  \frac{2+\bar\theta}{(6+\bar\theta)(3+\bar\theta)(1+\bar\theta )^2} +
  \mathcal O(\alpha^2)
\end{align*}
In case (ii),~\eqref{eq:distbasic} changes to
\begin{align*}
  \frac{d}{dt} \mathbf E_{N,\alpha}[e^{-\lambda R_{12}}] & = -\lambda
  \mathbf E_{N,\alpha}[e^{-\lambda R_{12}}] + 1 - \mathbf
  E_{N,\alpha}[e^{-\lambda R_{12}}] + \alpha |\beta| \mathbf
  E_{N,\alpha}[1_{Z_3=B} (e^{-\lambda R_{12}} - e^{-\lambda R_{23}})]
  \\ & \qquad \qquad \qquad \qquad \qquad +
  \alpha|\gamma| \mathbf E_{N,\alpha} [1_{Z_3=A}1_{Z_4=A} (e^{-\lambda
    R_{23}} - e^{-\lambda R_{12}})] + \mathcal O(1/N).
\end{align*}
As above, the term behind $\alpha|\beta|$ vanishes. Moreover,
\begin{align*}
  \mathbf E_{N,\alpha} [1_{Z_2=A}1_{Z_3=A} e^{-\lambda R_{12}}] & =
  \mathbf E_{N,\alpha} [1_{Z_2=A} e^{-\lambda R_{12}}] -
  \mathbf E_{N,\alpha} [1_{Z_2=A}1_{Z_3=B} e^{-\lambda R_{12}}],\\
  \mathbf E_{N,\alpha} [1_{Z_3=A}1_{Z_4=A} e^{-\lambda R_{12}}] & =
  \mathbf E_{N,\alpha} [1_{Z_4=A} e^{-\lambda R_{12}}] - \mathbf
  E_{N,\alpha} [1_{Z_3=B} 1_{Z_4=A}e^{-\lambda R_{12}}].
\end{align*}
Therefore, in equilibrium, 
\begin{align*}
  \mathbf E_{N,\alpha} [e^{-\lambda R_{12}}] & = \frac{1}{1+\lambda} +
  \frac{\alpha|\gamma|}{1+\lambda} \mathbf
  E_{N,0}[1_{Z_2=A}1_{Z_3=A}e^{-\lambda
    R_{12}}-1_{Z_3=A}1_{Z_4=A}e^{-\lambda R_{12}}] + \mathcal
  O(\alpha^2, 1/N) \\ & = \frac{1}{1+\lambda} - \frac{\alpha
    |\gamma|}{1+\lambda} \mathbf E_{N,0}[1_{Z_2=A}1_{Z_3=B}e^{-\lambda
    R_{12}}-1_{Z_3=A}1_{Z_4=B}e^{-\lambda R_{12}}] + \mathcal
  O(\alpha^2, 1/N).
\end{align*}
Using the result from case (i), we are done with case (ii). Cases
(iii) and (iv) are similar, because the term behind $\alpha|\beta|$
vanishes. This finishes the proof.


\end{document}